\documentclass[epj,twocolumn]{webofc}

\usepackage[varg]{txfonts}   % Web of Conferences font
\woctitle{The Innermost Regions of Relativistic Jets and Their Magnetic Fields}
\begin{document}
\title{A sensitive study of the peculiar jet structure HST-1 in M87}
%
% subtitle is optional

\author{Carolina Casadio\inst{1}\fnsep\thanks{\email{casadio@iaa.es}},
       Jos\'{e} L. G\'{o}mez\inst{1},
        Marcello Giroletti \inst{2},
        Gabriele Giovannini\inst{2,3},  
        Kazuhiro Hada\inst{2,4,5},
        Christian Fromm\inst{6},
        Manel Perucho\inst{7}, \and
        Jos\'{e}-Mar\'{i}a Mart\'{i}\inst{7}
                % etc.
}

\institute{Instituto de Astrof\'{i}sica de Andaluc\'{i}a, CSIC, Apartado 3004, 18080 Granada, Spain
\and
           INAF Istituto di Radioastronomia, via Gobetti 101, 40129 Bologna, Italy
 \and
           Dipartimento di Astronomia, via Ranzani 1, 40127 Bologna, Italy
 \and
          Department of Astronomical Science, The Graduate University for Advanced Studies (SOKENDAI), 2-21-1 Osawa, Mitaka, Tokyo 181-8588, Japan           
 \and         
          National Astronomical Observatory of Japan, Osawa, Mitaka, Tokyo 181-8588, Japan
 \and
 	  Max-Planck-Institut f\"ur Radioastronomie, Auf dem H\"ugel 69, 53121 Bonn, Germany
 \and
          Departament d'Astronomia i Astrofisica, Universitat de Val\`encia, Dr. Moliner 50, 46100 Burjassot, Valencia, Spain        
          }

\abstract{%
  To obtain a better understanding of the location and mechanisms for the production of the gamma-ray emission in jets of AGN we present a detailed study of the HST-1 structure, 0.8 arcsec downstream the jet of M87, previously identified as a possible candidate for TeV emission. HST-1 shows a very peculiar structure with superluminal as well as possible stationary sub-components, and appears to be located in the transition from a parabolic to a conical jet shape, presumably leading to the formation of a recollimation shock. This scenario is supported by our new RHD simulations in which the interaction of a moving component with a recollimation shock leads to the appearance of a new superluminal component. To discern whether HST-1 is produced by a recollimation shock or some other MHD instability, we present new polarimetric 2.2 and 5 GHz VLBA, as well as 15, 22 and 43 GHz JVLA observations obtained between November 2012 and March 2013.  
}
\maketitle
\section{Introduction}
In the era of the Fermi satellite and of the new generation Cherenkov telescopes, there is an active debate over the location and the mechanisms for the production of MeV to very high energy (VHE) gamma rays in AGN jets. M87 is a privileged laboratory for a detailed study of the properties of jets, owing to its conspicuous emission at all wavelengths and the combination of proximity (D=16 Mpc) and massive black hole (MBH $\sim6.4\times10^9$ $M_{\odot}$), resulting in a scale of 1 mas $\sim150$ $R_{s}$.
Despite many observational and theoretical efforts \cite{Aharonian2006,Stawarz2006,Cheung2007,Acciari2009}, the location of the VHE emission site in M87 is still elusive. The inner jet region is favored by the short TeV variability timescales and the simultaneous increase in radio flux density for the VHE event in 2008. The presence of compact substructures with superluminal motion and the simultaneous radio and X-ray flare during the 2005 event favor the location at the HST-1 complex about 0.8'' downstream the jet. Finally, the lack of prominent low energy flares in both the core and the HST-1 region associated to the 2010 TeV event further complicates the scenario \cite{Abramowski2012}.

\begin{figure}[t]
\centering
\includegraphics[scale=0.54,clip]{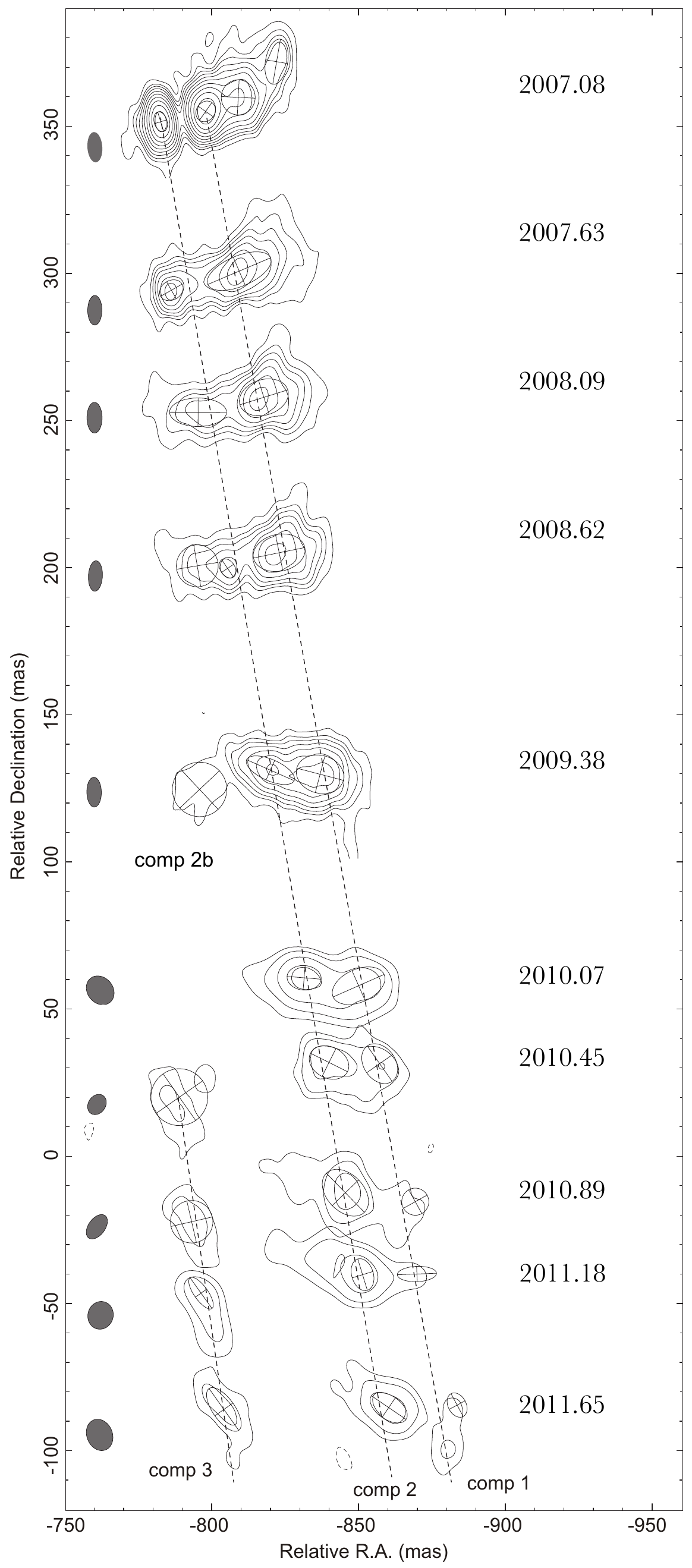}
\caption{HST-1 contour plots at 1.7 GHz (VLBA) from 2007.8 to 2009.38 and 5 GHz (EVN)  from 2010.07 to 2011.65. Model fit components are overlaid to the images. The contour plots are spaced vertically proportionally to the time interval between the relative epochs. The axis represent the relative (RA, Dec) coordinates from the core for the first image. From Giroletti et al.~\cite{Giroletti2012}.}
\label{M87-Gi}       % Give a unique label
\end{figure}

\begin{figure}[t]
\centering
\includegraphics[scale=0.31,clip]{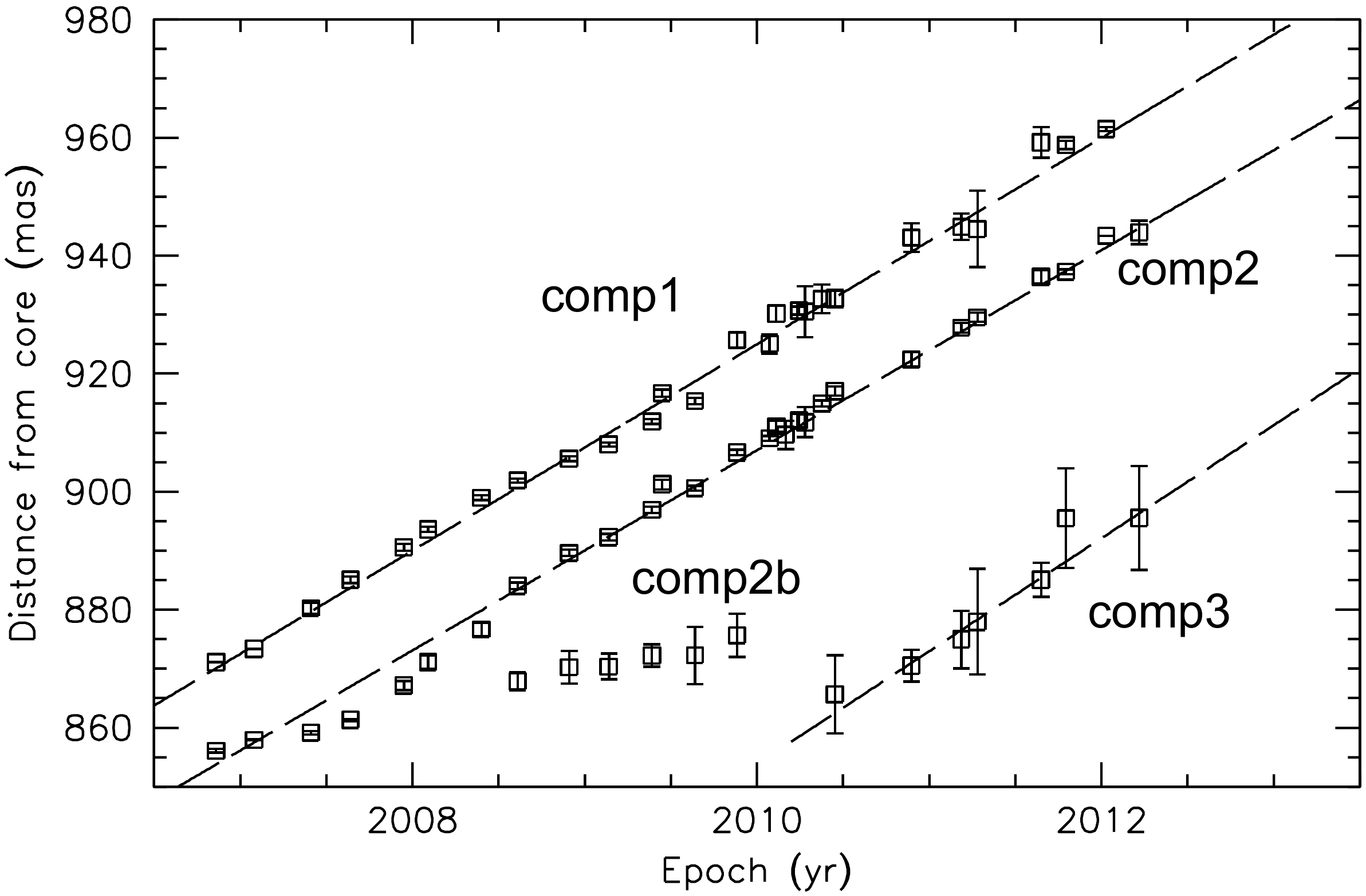}
\caption{Distance of compact components within HST-1 as a function of time. Note the new component, {\it comp3}, appearing in late 2010. From Giroletti et al.~\cite{Giroletti2012}.}
\label{Comp-Gi}       % Give a unique label
\end{figure}

Giroletti et al.~\cite{Giroletti2012}, presents a monitoring combining new e-EVN data and archival VLBA images for a total of 24 observations between November 2006 and October 2011, providing a detailed analysis of the radio images (see also Figs.~\ref{M87-Gi} and \ref{Comp-Gi}). The main results obtained can be summarize as follows: i) HST-1 in the radio band has a size of $\geq$100 mas, with sub-structures on smaller angular scales; i) two sub-components within HST-1 ($\it{comp1}$ and $\it{comp2}$ in Figs.~\ref{M87-Gi} and~\ref{Comp-Gi}) are moving regularly with superluminal velocity, having shown a displacement of more than 80 mas between 2006.9 and 2011.8 (v=4$\it{c}$); iii) a slower moving component ($\it{comp2b}$ in Figs.~\ref{M87-Gi} and~\ref{Comp-Gi}) was detected between 2008.5 and 2010 on the wake of the superluminal component $\it{comp2}$, in a similar fashion as the "trailing" components studied by numerical simulations \cite{Agudo2001} and observed in other sources \cite{Jorstad2005}; iv) in the latest EVN epochs (after 2010) a new substructure ($\it{comp3}$ in Figs.~\ref{M87-Gi} and~\ref{Comp-Gi}) has appeared at about 875 mas from the core, upstream of components $\it{comp1}$ and $\it{comp2}$, and in a position similar to that of the previous component $\it{comp 2b}$.

According to Stawarz et al.~\cite{Stawarz2006} the region at about 800--900 mas from the core is peculiar because at this position a reconfinement nozzle is expected due to the interaction of the jet with the external medium. The possibility that HST-1 corresponds to a recollimation shock is also supported by semi-analytical and numerical MHD models \cite{Gracia2009,Bromberg2009,Nalewajko2012} and by recent VLBI observations showing that HST-1 corresponds to the location at which the jet of M87 changes from a parabolic to a conical shape \cite{Asada2012}.

There are, however, no clear observational indications for the existence of a stationary feature in HST-1 associated with the recollimation shock --- as expected from numerical 
simulations \cite{Gomez1995,Gomez1997} and observed in other sources \cite{Agudo2012} ---, other than the earlier 1.7 GHz VLBA observations by Cheung, Harris, \& Stawarz \cite{Cheung2007}, which measured an upper-limit for the proper motion of the upstream region of HST-1 (labeled by these authors as HST-1d) of 0.25$\it{c}$. 
\label{intro}
\section{Relativistic hydrodynamic and emission simulations}
\label{sec-1}
\begin{figure*}[t]
\centering
\includegraphics[scale=0.68,clip]{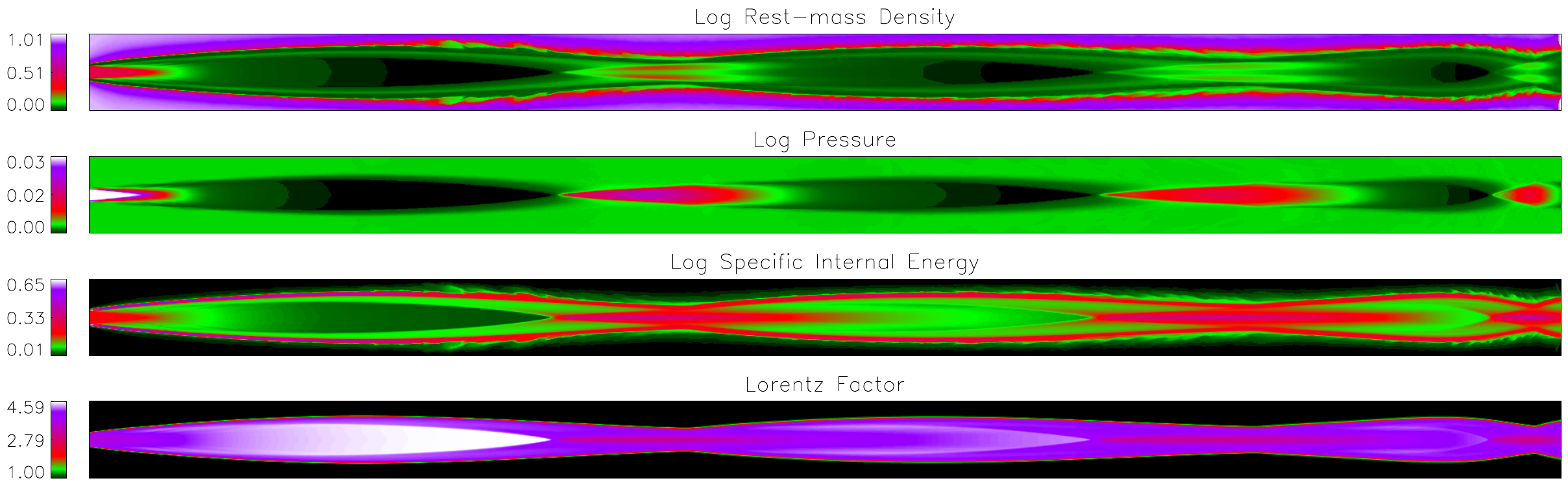}
\caption{Relativistic hydrodynamical simulation of a jet with an initial over-pressure 10 times larger than the external medium, leading to a set of recollimation shocks.}
\label{sims_st}       
\end{figure*}

\begin{figure*}[t]
\centering
\includegraphics[scale=0.68,clip]{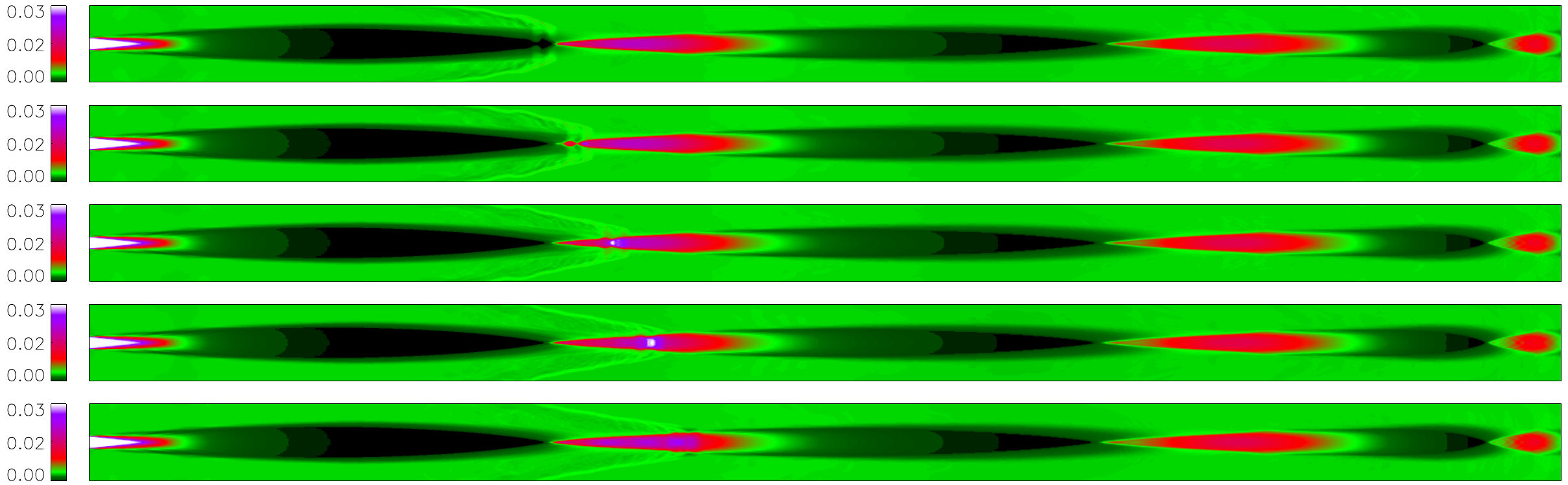}
\caption{Five snap-shots in the time evolution (from top to bottom) of the jet particle pressure after the introduction of a short increase in the jet inlet pressure by a factor of 8. The perturbation interacts with the second recollimation shock leading to a significant increase in pressure.}
\label{sims_HD}
\end{figure*}

\begin{figure}[t]
\centering
\includegraphics[scale=0.67,clip]{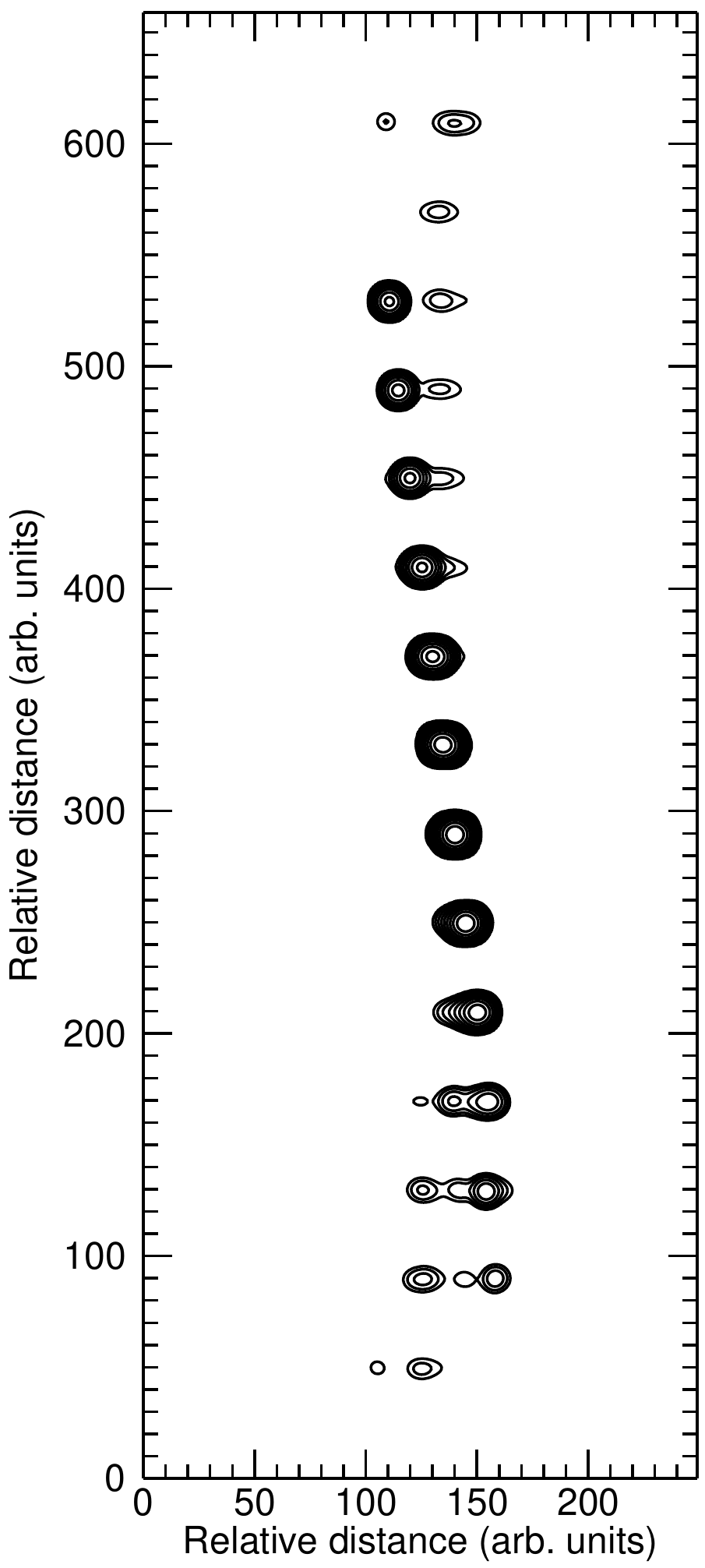}
\caption{Fifteen snap-shots in the time evolution (from top to bottom) of the simulated synchrotron emission maps obtained using the relativistic hydrodynamical models shown in Figs. 3 and 4 as input. Only the region corresponding to the interaction of the moving component with the recollimation shock is shown for clarity. Eleven contours in factors of $1/2^{n}$ (with n=0 to 10) of the absolute peak-flux are plotted. Note that the passage of the perturbation through the recollimation shock leads to the appearance of a new superluminal component.}
\label{sims}      
\end{figure}

To obtain a better understanding of the jet dynamics associated with the HST-1 region in M87 we have performed relativistic hydrodynamical simulations using the numerical finite-volume code $\it{Ratpenat}$, which solves the equations of relativistic hydrodynamics in conservation form using high-resolution-shock-capturing methods (\cite{Perucho2010}, and references therein). In order to obtain a series of strong recollimation shocks the jet is launched with an initial over-pressure 10 times larger than the external medium. Figure~\ref{sims_st} shows the rest-mass density, pressure, specific internal energy, and Lorentz factor for the stationary model. A perturbation in the jet inlet, consisting of a brief increase in the injection pressure by a factor of 8, quickly develops into a moving shock that later on interacts with the second recollimation shock, leading to a significant increase in its pressure (see Fig.~\ref{sims_HD}).
Figure~\ref{sims} shows a sequence of 15 total intensity images obtained by computing the synchrotron emission using the hydrodynamical results as input (for details of the numerical model used see \cite{Gomez1995},\cite{Gomez1997},\cite{Agudo2001},\cite{Aloy2003},\cite{Mimica2009}). A magnetic field in equipartition with the particle energy density and oriented predominantly in the direction of the jet bulk flow is assumed. Note that only the region corresponding to the interaction of the moving component and the recollimation shock is shown in Fig.~\ref{sims} for clarity. The interaction of the moving component with the recollimation shock leads to a significant increase in the particle and magnetic field energy density (see also the study of a possible shock-shock interaction in CTA~102 \cite{Fromm11,Fromm13}), leading to the sudden appearance of a new superluminal component in the emission maps by the third epoch shown of Fig.~\ref{sims}.
The appearance of the new superluminal component due to the interaction with the recollimation shock is in good agreement with our observations of M87 shown in Fig.~\ref{M87-Gi}, where component $\it{comp3}$ suddenly appears in a position similar to that observed previously for $\it{comp2b}$. Our simulated images of Fig.~\ref{sims} show that the quasi-stationary component associated with the recollimation shock (visible only in the first five and last four epochs shown in Fig.~\ref{sims}) is significantly dimmer than the moving component and its position can shift as a consequence of the passing of the superluminal component, as seen in previous numerical simulations \cite{Gomez1997}.
\section{Observations and data reduction}
\label{sec-2}
To test our hypothesis of a recollimation shock in HST-1 region we have performed new polarimetric observations of the radiogalaxy M87 with the VLBA at 2.2 and 5 GHz, and with the JVLA at 15, 22 and 43 GHz in A-configuration. Our polarimetric multi-frequency study is also aimed to probe the polarized emission structure along the jet and any possible Faraday rotation. The VLBA array at these two frequencies provides the resolution and the sensitivity needed to study in detail the structures in HST-1 and to follow the kinematics and flux density evolution of subcomponents in HST-1. The program consists of three epochs separated by 6 months each, of which here we present the observations of the first epoch (2013 March 9). The JVLA observations consist of 7 observing blocks between 2012 October 28 and 2012 December 22 of approximately three hours each.
\subsection{VLBA data}
\begin{figure*} 
\centering
\vspace*{0.01cm}
\includegraphics[scale=0.45,clip]{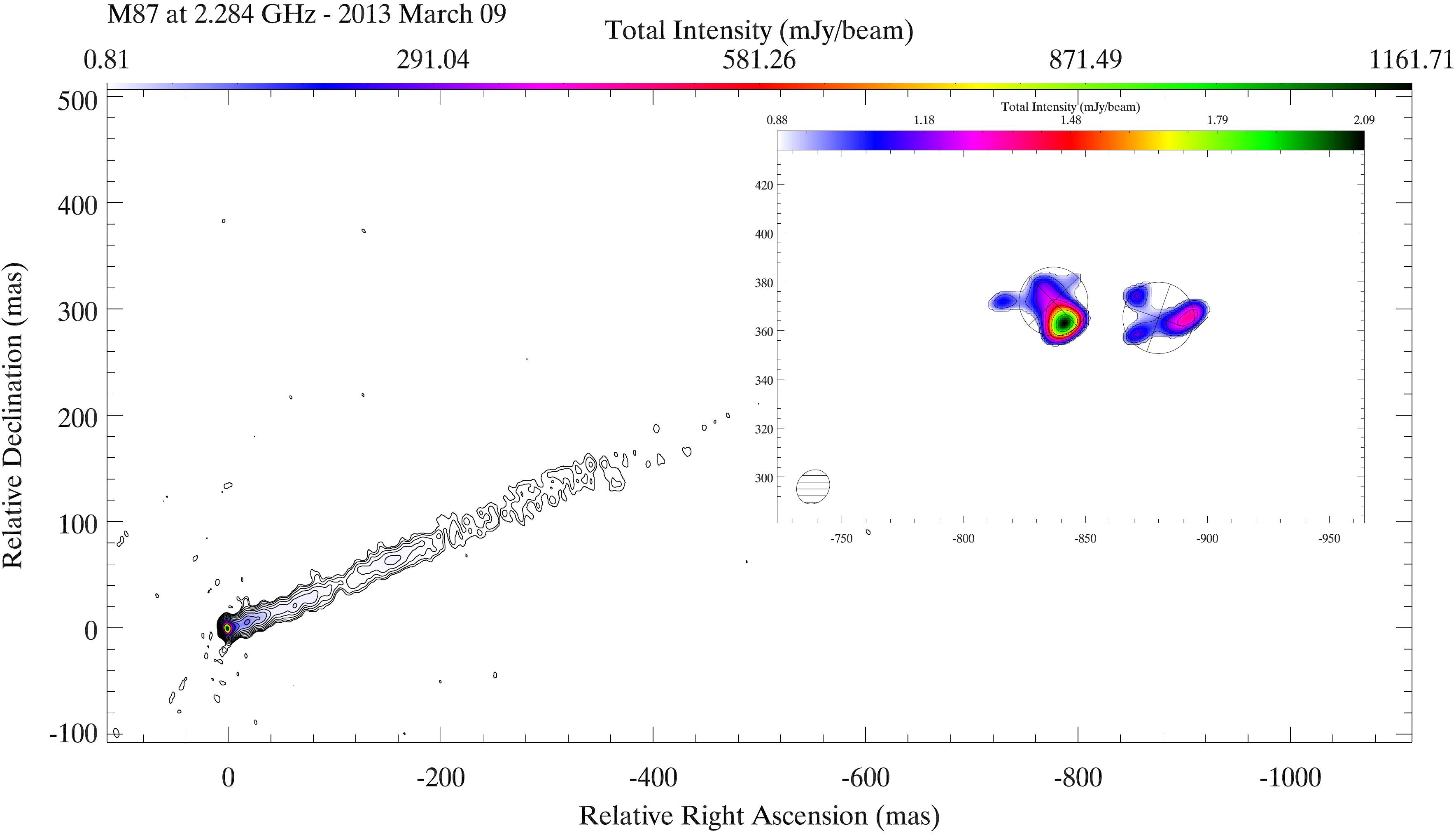}
\caption{VLBA observation at 2.2 GHz of M87 in 2013 March 09. Total intensity is shown in contours and colors. Contours are traced at 0.07, 0.13, 0.23, 0.42, 0.76, 1.38, 2.51, 4.56, 8.28, 15.03, 27.29, 49.56 and 90$\%$ of the peak brightness of 1.16 Jy/beam. The restoring beam is 8.8$\times$4.61 mas at 15.8$^{\circ}$. The panel inside the figure shows a zoom of the HST-1 region, obtained with a taper and resulting in a beam of 14.66$\times$12.85 mas at -38.2$^{\circ}$. Modelfits components in HST-1 are overlaid to the image and contours plot, where contours are traced at 38.3, 44.2, 50.9, 58.7, 67.7, 78.1, 90$\%$ of the peak brightness of 2.1 mJy/beam.  }
\label{HST1-zoom}      
\end{figure*}

\begin{figure*} 
\centering
\includegraphics[scale=0.45,clip]{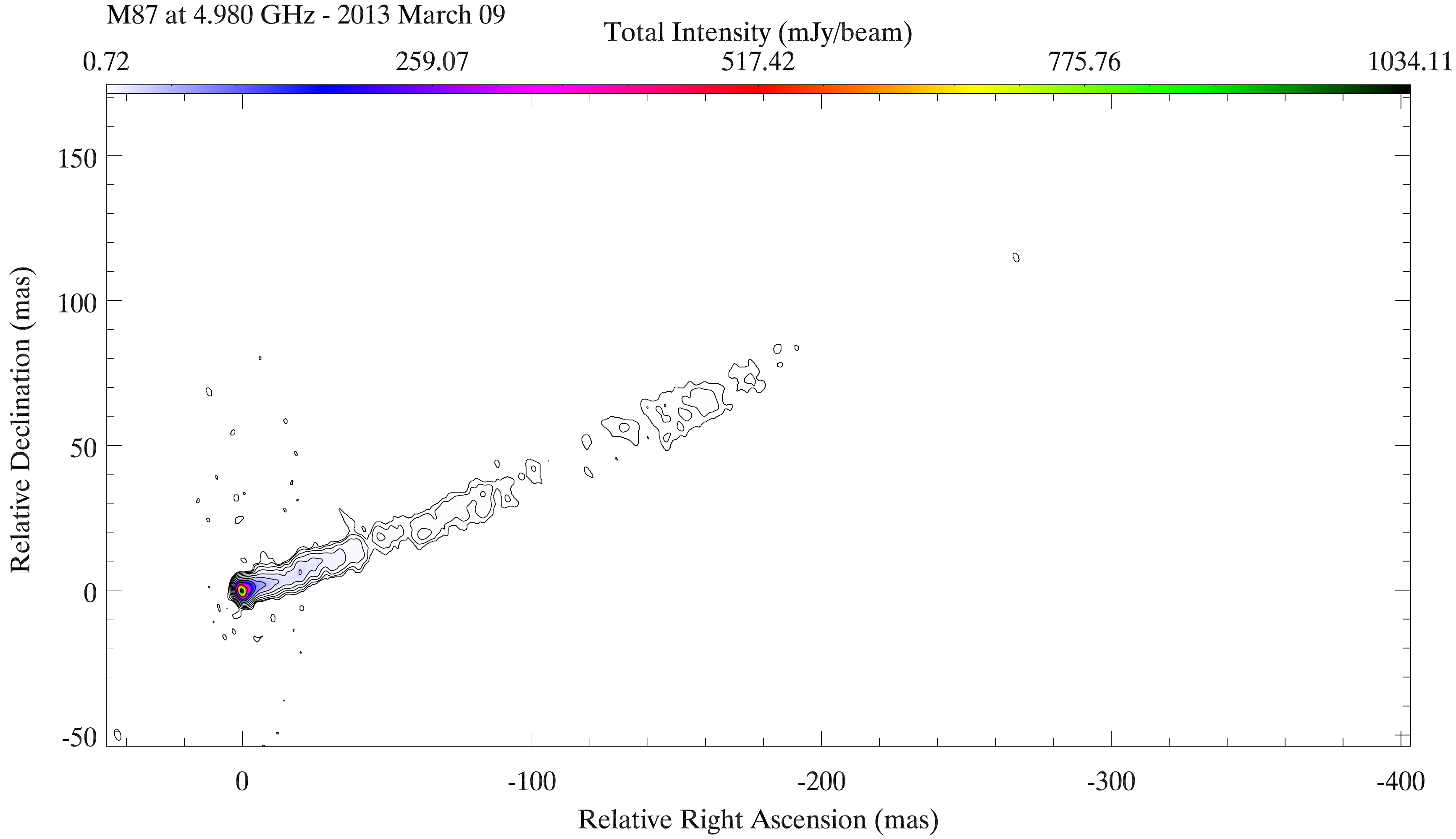}
\caption{VLBA observation at 5 GHz of M87 in 2013 March 09. Total intensity is shown in contours and colors. Contours are traced at 0.07, 0.14, 0.29, 0.6, 1.23, 2.51, 2.14, 10.51, 21.5, 43.99, 90$\%$ of the peak brightness of 1.034 Jy/beam. The restoring beam is 3.95$\times$2.06 mas at 18.03$^{\circ}$. }
\label{M87_5GHz}      
\end{figure*}

The calibration and imaging of the data was performed using AIPS and Difmap packages in the usual manner \cite{Leppanen1995}. Figures \ref{HST1-zoom} and \ref{M87_5GHz} show the VLBA images obtained in 2013 March 9 at 2.2 and 5 GHz, respectively. For both images we obtain a sensitivity of about 0.1--0.2 mJy/beam. A gaussian taper to the visibility data at 2.2 GHz has been applied in order to enhance the emission region of the HST-1 complex. The resulting image is shown in the inset panel of Fig.~\ref{HST1-zoom}, together with a fit of the visibilities to circular gaussian models to describe the emission structure in HST-1.

\subsection{JVLA data}
\begin{figure*}
\centering
\includegraphics[scale=0.82,clip]{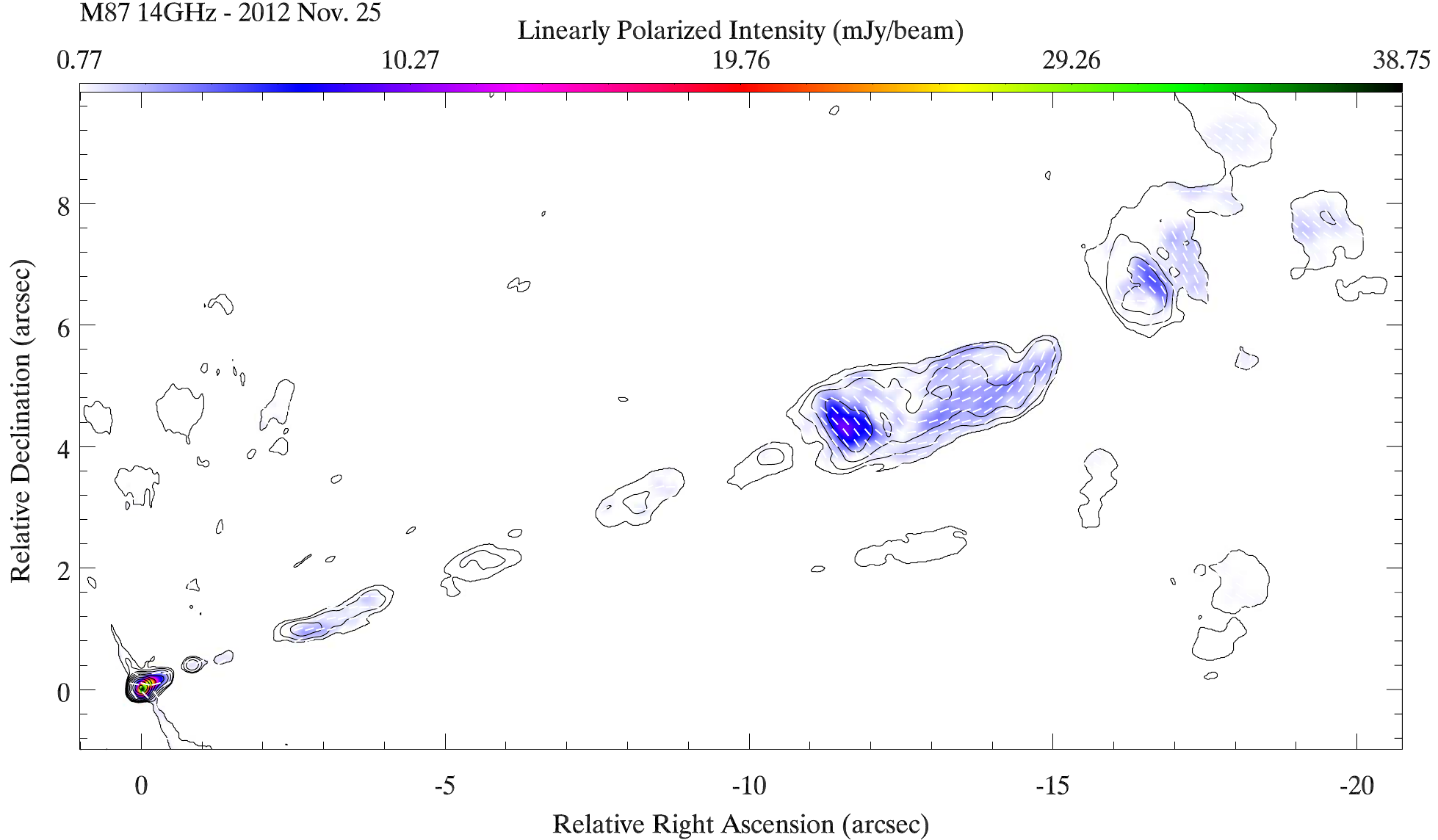}
%\vspace*{0.01cm}
\caption{JVLA observation at 15 GHz of M87 in 2013 November 25. Total intensity is shown in contours and polarized flux in colors. White bars, of unit length, indicate the magnetic field direction (uncorrected for Faraday rotation). Contours are traced at 0.04, 0.09, 0.22, 0.52, 1.24, 2.91, 6.87, 16.19, 19.38, 17.90$\%$ of the peak brightness of 2.93 Jy/beam. The restoring beam is 0.17$\times$0.165'' at -47$^{\circ}$.}
\label{M87_JVLA}      
\end{figure*}

Analysis of the JVLA data has been carried out with CASA following the standard procedure described in the CASA cookbook. Calibration of the bandpass was obtained through observations of 3C~279, while 3C~286 was used as flux density calibrator and to determine the instrumental polarization and absolute orientation of the electric vectors position angle. A preliminary image at 15 GHz, obtained with a total on source integration time of approximately 12 minutes, is shown in Fig.~\ref{M87_JVLA} for the observations carried out on 2012 November 25. Analysis of the emission structure and Faraday rotation detected in the different knots, including HST-1, will be presented in a forthcoming paper.

\section{Results}
Our VLBA observations reveal a signifiant decrease in the flux density of HST-1 as compared with previous epochs \cite{Giroletti2012}. At 2.2 GHz our VLBA image shows a peak brightness for HST-1 of 2.1 mJy/beam. No detection of HST-1 was obtained for the 5 GHz VLBA, but we can estimate an upper limit for the peak flux density at 5 GHz of 0.5 mJy/beam (5-$\sigma$).

\begin{figure}[t]
\centering
\includegraphics[scale=0.3,clip]{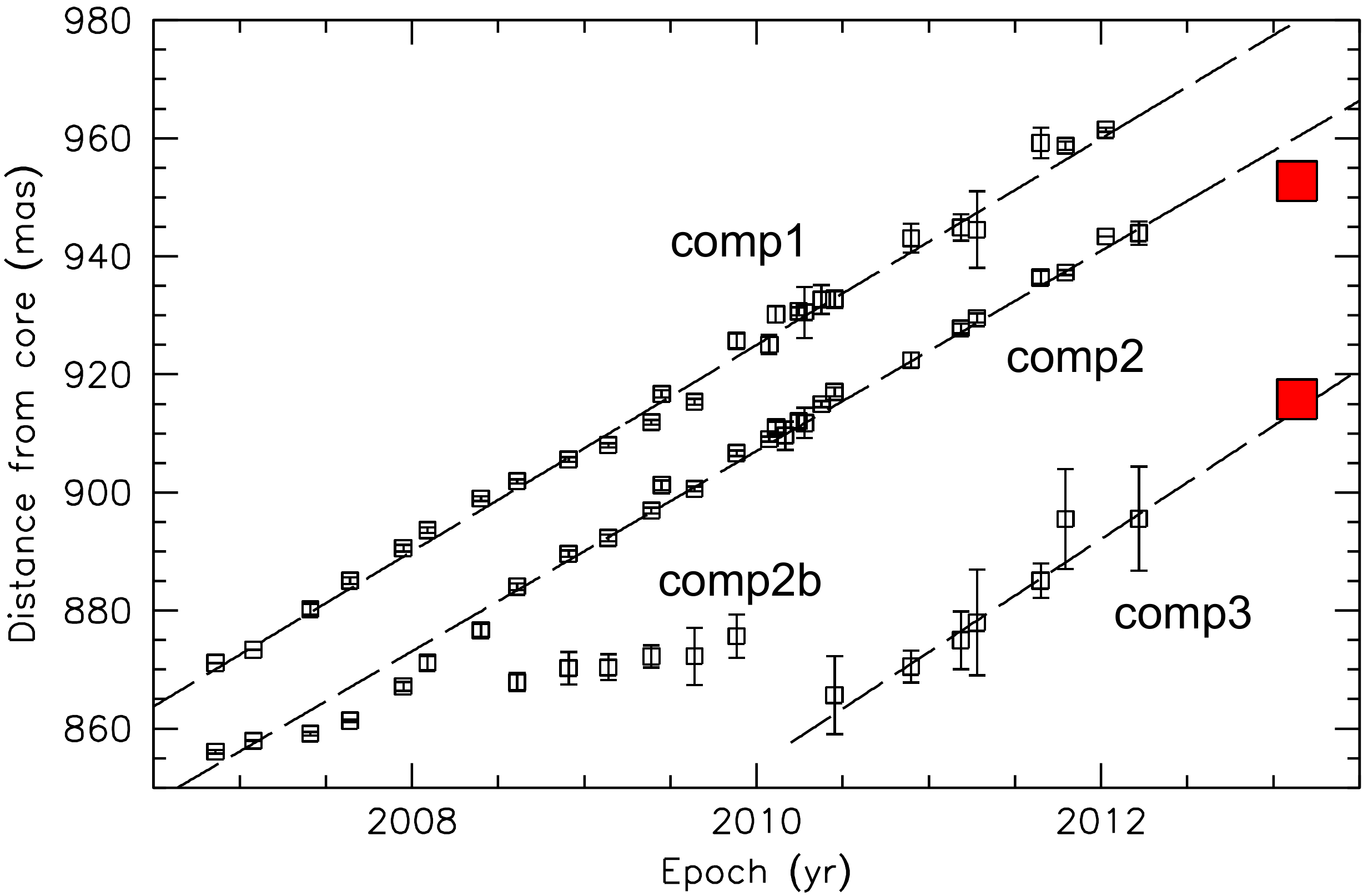}
\caption{Distances of compact components in HST-1 as a function of time as registered in Giroletti et al.~\cite{Giroletti2012}. The two red squares represent the position of our two Gaussian modelfit components detected in HST-1 in the image at 2.2 GHz with the VLBA on 2013 Marth 9 (Fig.~\ref{HST1-zoom}).}
\label{fig-9}      
\end{figure}

  The HST-1 region at 2.2 GHz can be model fitted into two different components, one located at $\sim$913 mas from the core and a second one at $\sim$953 mas. Comparison of the position of these two components with those found by Giroletti et al.~\cite{Giroletti2012} reveals that we can associate these components to {\it comp3} and {\it comp2}, respectively, as shown in Fig.~\ref{fig-9}. In our new observations component {\it comp3} is brighter than {\it comp2}, which together with the lack of detection of {\it comp1} suggests that subcomponents in the HST-1 region decrease in flux density as they move downstream, probably due to adiabatic expansion. 
  
  Comparison of our JVLA data with previous VLA archive data also shows a significant decrease in flux density in HST-1. Our 2012 November 25 observations at 15 GHz show a total flux of $\sim$9 mJy, while it was observed to be $\sim$27 mJy and $\sim$52 mJy in 2003 and 2006, respectively. The peak flux density of the core has however increased from $\sim1.9$ Jy/beam in 2006 to $\sim2.9$ Jy/beam in 2012.
  
  Posterior analysis of the polarization combining the VLBA and JVLA data should provide further information regarding the nature of the HST-1 region, and whether it indeed corresponds to a recollimation shock in the jet, as suggested by our RHD and emission simulations. We also expect that our ongoing VLBA monitoring will reveal a new brighting of the HST-1 region, providing a clear detection with the enhanced resolution at 5 GHz.
  
  This research has been supported by the Spanish Ministry of Economy and Competitiveness grants AYA2010-14844 and AYA2010-21097-C03-01, and by the Regional Government of Andaluc\'{\i}a (Spain) grant P09-FQM-4784. The VLBA is an instrument of the National Radio Astronomy Observatory, a facility of the National Science Foundation operated under cooperative agreement by Associated Universities, Inc.


\end{document}